\documentclass[a4paper,10pt]{article}
\usepackage{amsmath,amssymb,amscd,amsthm}
\usepackage[all]{xy}
\begin{document}
\title{Dark Matter and Torsion}

\author{G. Grensing\\ Institut f\H ur Theoretische Physik \\und Astrophysik \\ 
Universit\H at Kiel \\D-24118 Kiel, 
Germany\\ \\g.grensing@theo-physik.uni-kiel.de}

\date{December 2019}

\maketitle

\begin{abstract}
Superheavy right-handed Majorana neutrinos are proposed as a promising candidate for dark matter, with dynamical axial torsion as the mediating agent.
\end{abstract}
\numberwithin{equation}{section}
\vspace{1.5cm}

Although almost a whole century has passed since its perception, the nature and origin of dark, i.e. non-luminous,
 matter (DM) still remains a mystery. A plethora of probable candidates is available; the bestiary contains neutralinos, axions, massive astrophysical compact halo 
objects  (MACHOs), macros, sterile neutrinos, gravitinos, weakly interacting massive particles (WIMPs), and the like. The currently favoured aspirants are the latter ones, but neither the mass nor the spin of these WIMPs is known (see $\left[1\right]$ for  reviews). Restricting ourselves to particle DM, in spite of extensive searches at colliders and  underground experiments in the eV and GeV mass-range, no sign of a WIMP has as yet been detected; the search was even extended to the TeV scale, with the same outcome $\left[2\right]$.  

However, some new aspects of the problem have evolved in the cosmological  context. From recent work on the issue of  
matter-antimatter asymmetry $\left[3\right]$ it has emerged that its explanation requires an intricate analysis of leptogenesis, which in turn requires that the mass of DM particles is roughly restricted to the range $10^{10} - 10^{15}$ GeV, saying that they must be extremely heavy $\left[4\right]$. Similar energy scales have also proven to be necessary in order to comprehend the production mechanism of such super-heavy DM particles $\left[5\right]$, being dubbed WIMPzillas. They must have been produced at an early epoch of the history of the universe, through particle creation at the end of inflation, in order to generate the measured DM abundance; from astronomical and cosmological data this is known to account for $25 \%$ of the matter-energy content of the universe. The production of WIMPzillas functions through the so-called freeze-in mechanism $\left[6\right]$ and requires energies of the order of the reheating temperature, i.e. $10^{13}$ GeV, and this scale is indeed within the range of masses as announced. Further support for such superheavy DM particles comes from the observation of exceptional air showers with energies above  $10^{11}$ GeV $\left[7\right]$, for which the above scenario could offer a natural explanation $\left[8\right]$. 

In the present paper we suggest, as here and there  already noticed in the literature, that the right-handed Majorana neutrinos of the 
see-saw mechanism $\left[9\right]$ can serve as a viable candidate for dark matter; what is more, we even claim these right-handed fermions to constitute the natural extension of the standard model (SM) of particle physics, providing the new scale of the physics beyond the standard model. Due to the Majorana property there is no coupling to the electromagnetic field; if we even require that they are blind to the electroweak forces, then they belong to the WIMPzilla category of cold dark matter (CDM), and as such our choice is strongly supported by the results just having been reported in the precedinng paragraph. We are thus motivated to identify dark matter with right-handed Majorana fermions. In order to get access to the mediator field for DM particles, we need an extension of general relativity (GR). We choose the well known one by torsion of \'{E}. Cartan 
$\left[10\right]$, which in the present setting will find its natural physical realization. As only the axial part of torsion couples to 
(Majorana) fermions, the interaction with matter fields, i.e. non-gauge fields, is thus fixed. It remains to design the action for the torsion field. What comes to mind is to resort to the Einstein-Cartan theory (ECT), in which the Riemannian curvature scalar is simply replaced by the corresponding scalar of a Riemann-Cartan space; but then torsion would not be a dynamical agent. At this decisive point at least one should esteem the spectral action principle of Alain Connes et al. from 1997 $\left[11\right]$; as of that year the rather unique option is available to determine a gauge field action within the framework of  non-commutative geometry  
(NCG) from first principles. The result, torsion is an extremely massive, dynamical axial vector field;  it couples directly to the Higgs field. Hence, noncommutative geometry also predicts what is known in the literature as the Higgs portal $\left[1,12\right]$.
As a last issue, one has to guarantee that the evolution of the universe is not essentially altered, i.e. the Friedmann equations of 
cosmology must remain intact; this point is discussed in the last section where it is shown that, due to the DM property of being made up of Majorana fermions, such superheavy particles can indeed account for the relic density of dark matter at present.

\section{Majorana fermions and dark matter}

The hierarchy and triviality problem $\left[13\right]$ of the standard model indicate that one should expect new physics above the electroweak energy scale of $\Lambda_{\text{EW}} \approx 300$ GeV. Hence the standard model has to be viewed as an effective theory; this means that the Lagrangian of the theory beyond the SM (BSM) can be expanded in terms of an ultraviolet cutoff $M$; for energies below this scale the BSM is valid. The first term of the expansion proportional to $M^0$ is the SM Lagrangian, and the next term 
proportional to $M^{-1}$ is a mass-dimension five term of the form
\begin{equation}\label{1,1}
\frac{(\bar{H}L)(\bar{H}L)}{M}
\end{equation}  
where $H$ denotes the Higgs- and $L$ the lepton-doublet; it is of non-renormalizable type. For the vacuum expectation value of the Higgs $\left<H\right>=v$ it takes the form .
\begin{equation}\label{1.2}
\frac{v^2(\bar{\nu}v)}{M}
\end{equation}  
which is the mass term for the three family neutrinos $\nu_i$. Indeed, there is confirmed evidence for neutrino masses; as known, none of these masses $m_{\nu_i}$ can exceed 
$0.1$ eV, a tiny value being far outside the the typical scales of the SM; also, at least one of these masses, denoted $m_{\nu}$, is nonzero. Hence the cutoff $M$ is of the order 
$M\approx 10^{14}$ GeV, roughly the unification scale, i.e. within the mass range of the WIMPzillas. Thus, since the neutrino masses can only be extremely small, as compared to the typical scales of the SM, one encounters a fine tuning problem, and this is resolved according to the above reasoning. These arguments can be put further $\left[9,14\right]$ on tentatively assigning to the cutoff M a particle, denoted by $N$, which turns out to be of Majorana type and is the (predominantly) right-handed partner of the left-handed 
$\nu$-neutrino. 

To resume, the experimental fact that at least one of the neutrinos must be massive is the first indication for the need to go beyond the standard model. We go even further and endow these superheavy right-handed Majorana neutrinos $N$ of mass $M$ with physical reality by boldly requiring that they constitute the new sector of the physics beyond the standard model, which in addition we identify with the dark matter sector. The Majorana property implies that $N$ is electrically neutral, i.e. invisible or dark.

Of course, the version of the see-saw mechanism as given is greatly simplified. In particular we have not mentioned that for each neutrino $\nu_i$ of the three fermion families there is a right-handed companion $N_i$ of mass $M_i$, and consequently there is a variety of possible see-saw mechanisms available  $\left[1\right]$. But we desist from these extensions since we only need the shortened version for the applications we have in mind. As a last point, further support for the need to introduce N-neutrinos with factual reality comes from Grand Unified Theories (GUT) $\left[15\right]$. For the group extension $SO(10)$ of $SU(3)\times SU(2)\times U(1)$ the unification of the SM predicts in addition to the three fermion sectors of quarks and leptons three singlets $N_i$ - as follows from the decomposition of the $16$-dimensional space of half-spinors of $SO(10)$ into $SU(5)$-invariant subspaces according to the decomposition
$(16)=(\bar{5})\oplus (10)\oplus (1)$, 
where the singlet $(1)$ corresponds to the $N$-neutrino.

\section{Coupling of dark matter to gravity}

Having identified the particle species that, at least predominantly, makes up the dark sector, i.e. the superheavy right-handed Majorana fermions or WIMPzillas, it remains to couple these particles to the gravitational field, and to specify the mediator field for the interaction between the dark matter particles.

We have achieved to establish the dark sector by means of a minimal extension of the SM. The same demand we impose on the present task, the search for a minimal extension of standard GR in order to incorporate the mediator sector.

Majorana particles are Dirac fermions, being invariant under the charge conjugation operator $\mathcal{C}$, that is 
$\mathcal{C}\psi=\psi$ where $\psi$ is a Dirac spinor. The inclusion
of the gravitational field in the Dirac operator is best described by the Kibble-Sciama-Utiyama gauge approach to gravity 
$\left[16\right]$ (for reviews see $\left[17\right]$); a further upgrade is reached on using so-called affine Cartan connections 
$\left[18\right]$ which - in some sense - help to clarify the role of local translations. The point of departure is the fact that the nonlinear Poincar\'{e} group of translations $a$ and Lorentz transformations $\Lambda$ can be represented as a special subgroup of the 
5-dimensional general linear group, namely
\begin{equation}\label{2.1}
(a,\Lambda)=\begin{pmatrix}\Lambda^{\,\alpha}{}_{\beta}&a^{\,\alpha}\\0&1\end{pmatrix}_{\alpha,\beta
=0,1,2,3}
\end{equation}
so that a liner matrix realization of the Lie algebra operators of translations $P_{\alpha}$ and Lorentz transformations 
$M_{\alpha\beta}$ is available with the standard commutation relations. Then one can pass to local paramaters $a^{\alpha}(x)$ of translations and $\alpha^{\gamma\delta}(x)$ of Lorentz transformations and introduce the linear sum
\begin{equation}\label{2.2}
\lambda=-a^{\alpha}\,P^{}_{\,\alpha}+\frac{1}{2}\,\alpha_{}^{\gamma\delta}\,M^{}_{\,\gamma\delta}
\end{equation}
so that the localized $(a,\Lambda)=\exp(-i\lambda)$ is available. In the Yang-Mills fashion we are then able to define the connections
\begin{equation}\label{2.3}
\xi^{}_{\mu}(x)=-e^{\alpha}{}_{\mu}(x)\,P^{}_{\,\alpha}+\frac{1}{2}\,\omega^{\gamma\delta}{}_{\mu}(x)
\,M^{}_{\,\gamma\delta}.
\end{equation}
Here the $\omega^{\gamma\delta}{}_{\mu}(x)$ denote the gauge fields for Lorentz transformations and the $e^{\alpha}{}_{\mu}(x)$ the 
quasi-gauge fields for translations, called the co-frame or inverse tetrad. Indices $\alpha, \beta, \ldots$ are frame indices, and 
$\mu, \nu, \ldots$ coordinate indices; the former are raised by means of the Kronecker delta, and the latter by means of the metric tensor $g_{\mu\nu}=\delta_{\alpha\beta}e^{\alpha}{}_{\mu}e^{\beta}{}_{\nu}$; frame indices can be transformed into coordinate indices by means of the frame and the coframe and vice versa. The transformation law of the gauge fields is now the standard one:
\begin{equation}\label{2.4}
(a,\Lambda)\cdot\xi_{\,\mu}=(a,\Lambda)(\xi_{\,\mu}+i\partial_{\,\mu})(a,\Lambda)^{-1}.
\end{equation}
From now on this is known territory; the simplest strategy is to pass to Lie algebra-valued differential forms 
$\xi=\xi^{}_{\mu}dx^{\mu}$; then the curvature 2-form is $\Xi=d\xi+\xi\wedge \xi$, with the explicit form
\begin{equation}\label{2.5}
\Xi=\frac{1}{2}\Big(-P_{\gamma}\,\Theta^{\gamma}{}_{\mu\nu}+\frac{1}{2}M_{\gamma\delta}\,\Omega^{\gamma\delta}{}_{\mu\nu}\Big)dx^{\mu}\wedge dx^{\nu}
\end{equation}
where
\begin{equation}\label{2.6}
\Theta^{\,\gamma}{}_{\mu\nu}=(\partial_{\mu}e^{\,\gamma}{}_{\nu}
-\partial_{\nu}e^{\,\gamma}{}_{\mu})-(\omega ^{\,\gamma}{}_{\mu\nu}-\omega^{\,\gamma}{}_{\nu\mu})
\end{equation}
are identified as the field stengths of translations, and   
\begin{equation}\label{2.7}
\Omega^{\,\gamma\delta}{}_{\mu\nu}=
(\partial_{\mu}\omega^{\,\gamma\delta}{}_{\nu}
-\partial_{\nu}\omega^{\,\gamma\delta}{}_{\mu})-\delta_{\,\alpha\beta}
(\omega^{\,\alpha\gamma}{}_{\mu}\omega^{\,\beta\delta}{}_{\nu}-
\omega^{\,\alpha\gamma}{}_{\nu}\omega^{\,\beta\delta}{}_{\mu})
\end{equation}
as the field strengths of rotations. In the traditional approach to differential geometry the latter one is related to the curvature tensor by
$R_{\,\mu\nu\rho}{}^{\tau}=-\Omega^{\,\tau}{}_{\rho\nu\mu}$ and the former to what is known as the torsion tensor through
$T_{\,\mu\nu}{}^{\,\rho}=-\Theta^{\,\rho}{}_{\,\nu\mu}$; the original definition of torsion is
$T_{\,\mu\nu}{}^{\,\rho}=\Gamma_{\,\mu\nu}{}^{\,\rho}-\Gamma_{\,\nu\mu}{}^{\,\rho}$ with the $\Gamma_{\,\mu\nu}{}^{\,\rho}$ the usual connection coefficients. Normally, torsion is required to vanish so that the connection coefficients $\Gamma_{\,\mu\nu}{}^{\,\rho}$ are symmetric in their lower indices and can solely expressed in terms of the metric tensor and its inverse. 

Here, however, torsion plays a central role and our starting point is the Dirac operator in spaces with torsion.
It is built from the covariant derivatives
\begin{equation}\label{2.8}
\nabla_{\mu}(\omega)=\partial_{\mu}-
\frac{i}{2}\,\Sigma_{\,\gamma\delta}\,\omega^{\,\gamma\delta}{}_{\mu}
\end{equation}
where the $\Sigma_{\,\alpha\beta}=(i/4)\left[\gamma^{\alpha},\gamma^{\beta}\right]$ are the generators of Lorentz transformations in the Dirac representation, and read
\begin{equation}\label{2.9}
D=i\gamma^{\alpha}e^{\mu}{}_{\alpha}\big(\nabla_{\mu}(\omega)-\frac{1}{2}\,\Theta_{\mu}\big)
\end{equation}
in which the last term built from the once contracted torsion $\Theta_{\mu}=\Theta^{\beta}{}_{\nu\mu}\,e^{\nu}{}_{\beta}$ is needed to make $D$ a (formally) selfadjoint operator

In order to isolate the torsion dependence, we use eq. \eqref{2.6} to express the rotational gauge fields 
$\omega^{\gamma\delta}{}_{\mu}$ in terms of (derivatives of) the tetrad, and torsion; one obtains
\begin{equation}\label{2.10}
\omega_{\rho\sigma\mu}=\overset{\scriptscriptstyle\circ}{\omega}_{\rho\sigma\mu}+\mathcal{K}_{\rho\sigma\mu}\,:\qquad\qquad
\mathcal{K}_{\rho\sigma\mu}=-\frac{1}{2}\big(\Theta_{\rho\sigma\mu}-
\Theta_{\mu\rho\sigma}+\Theta_{\sigma\mu\rho}\big)
\end{equation}
where the $\overset{\scriptscriptstyle\circ}{\omega}_{\rho\sigma\mu}$ denote the torsion-independent Ricci rotation coefficients.
Insertion into the Dirac operator gives
\begin{equation}\label{2.11}
D=\overset{\scriptscriptstyle\circ}{D}-\frac{i}{8}\,\gamma^{\alpha\beta\gamma}\Theta_{\alpha\beta\gamma}
\end{equation}
where $\gamma^{\,\alpha\beta\gamma}=\gamma^{\,[\alpha}\,\gamma^{\,\beta}\,\gamma^{\,\gamma]}$ and 
$\overset{\scriptscriptstyle\circ}{D}=
D|_{\Theta=0}=i\gamma^{\alpha}e^{\mu}{}_{\alpha}\overset{\scriptscriptstyle\circ}{\nabla}_{\mu}$ the Dirac operator for vanishing torsion. This form makes it manifest that the Dirac operator $D$ only depends on the totally skew-symmetric part of the torsion tensor.
Now restricting ourselves to 1+3 dimensions, we can pass to the axial torsion vector
\begin{equation}\label{2.12}
T^{\,5}{}_{\alpha}=\frac{1}{2}\,\varepsilon_{\alpha\beta\gamma\delta}\Theta^{\beta\gamma\delta}
\end{equation}
and using the identity
$\gamma^{\alpha\beta\gamma}=-\varepsilon^{\alpha\beta\gamma}{}_{\delta}\,\gamma^{\,5}\gamma^{\delta}$ one arrives at 
\begin{equation}\label{2.13}
D=i\gamma^{\alpha}\big(\overset{\scriptscriptstyle\circ}{\nabla}_{\alpha}+\frac{1}{4}\,\gamma^{}_{\,5}T^{\,5}{}_{\alpha}\big).
\end{equation}
which is the final form of the Dirac operator in spaces with torsion. Accordingly, the coupling to conventional gravity is standard, and the additional 
coupling to torsion is of an almost internal form since $\gamma^{\,5}$ commutes with the Lorentz generators.
 
We can return then to our central theme. According to the appointment having been made, the dark sector is defined as the space of fermions, denoted $\chi$ from now on, obeying $\mathcal{C}\chi=\chi$, i.e. each of its elements is mapped by the charge conjugation operator into itself. Since $\mathcal{C}$ commutes with local spin transformation, it is also invariant under local Lorentz transformations. Furthermore $D$ commutes with 
$\mathcal{C}$, and thus the Dirac operator  as well maps the dark sector into itself. These statements are all valid in the general case, with and without torsion. Therefore, from the explicit form
\eqref{2.13}  of the Dirac operator we read off that the axial torsion is to be identified as the mediating field for the dark matter sector. 

To recapitulate, in the preceding and the present section we have achieved to define the space of dark matter particles and their mutual interactions by means of a natural minimal extension of both the standard model and general relativity. In the latter case this insight came from the recognition that, in a certain sense, torsion  can be viewed as the curvature generated by local translations - the qualification 'in a certain sense` being used since in the flat case, i.e. for  vanishing gravitational fields, the tetrad does not 
vanish - as it is then equal to the unit matrix.

\section{Spectral action and torsion}

For the torsion extension of GR there is a venerable theory available, the Einstein-Cartan theory (ECT)(see $\left[17\right]$). It is obtained from the conventional Riemannian case on replacing the covariant derivative in the action for the Dirac field and the curvature term in the Einstein-Hilbert action by the corresponding expressions for nonzero torsion. The disadvantage of the EC theory is, torsion does not propagate since a kinetic term for torsion is absent. On the other hand, this disadvantage can be turned into a benefit in that torsion may be eliminated in favour of a long known 4-fermion 
self-interaction term, due to Utiyama and Kibble $\left[16\right]$ (see also $\left[19\right]$), which however is of non-renormalizable type. All in all, the ECT was studied over the years in all its facets; one might say that - adopting a saying - this theory is like a solution looking for a problem.

As the EC theory is of no use for the present purposes, it would amount to a really extreme advantage if a technique were available to compute, for a given set of gauge fields including gravity, the associated gauge field action; this then would replace the traditional trial and error method. Indeed, it exists and is due to Chamseddine and Connes  $\left[11\right]$. Their approach is called the spectral action principle (SAP) and relies on the properties of the square of the Dirac operator, which may be viewed as a kind of Hamilton operator in $n$ (even) space-time dimensions; it is understood that the transition from minkowskian to euclidean conventions was made. In order to cut off the spectrum of $D^2$ above a large but finite eigenvalue $\Lambda$, a characteristic function that can be disposed of is introduced, and in the form of certain moments 
$\chi^{}_{2k}$ it reappears in the final form of the spectral action: 
\begin{equation}\label{3.1}
S^{}_{\Lambda}=\sum_{k=0}^{n/2}\chi^{}_{2k}\,\Lambda^{n-2k}\frac{1}{(4\pi)^{n/2}}\int\text{tr}\,[a_{k}(D^2)]\sqrt{|g|}\,d^nx.
\end{equation}
The coefficients $[a_{k}(D^2)]$ come from the expansion of the heat-kernel operator
\begin{equation}\label{3.2}
\text{Tr}\,e^{-\beta D^2}=\frac{1}{(4\pi\beta)^{n/2}}\sum_{k\geq 0}\beta^{k}\int\text{tr}\,[a_{k}(D^2)]\sqrt{|g|}\,d^nx.
\end{equation}
In the Riemannian case these coefficients are known 
$\left[20\right]$, and also 
their generalization to nonzero torsion $\left[21\right]$. If the SAP is applied to the standard case of a Dirac operator without torsion and in the presence of further internal gauge fields, the well-known result with the cosmological constant, the scalar curvature, and the Maxwell term follows, amended by the square of the Weyl tensor. It is really  remarkable that, rescaling the prefactors in order to introduce conventional notation, they all carry the correct sign in front.  

The entire capability of the SAP only shows up, however, if one wants to include the Higgs field. In that case the whole apparatus of noncommutative geometry is needed. The essential notion of NCG is a so-called spectral triple $(\mathcal{A},\mathcal{H},D)$; in the standard case considered up to now it is built from a Dirac operator $D$, acting on the Hilbert space $\mathcal{H}$ of square-integrable spinors over a manifold $M$, and $\mathcal{A}$ is the commutative algebra of smooth functions on $M$, acting on $\mathcal{H}$ by pointwise multiplication. In the general noncommutative case the idea of an underlying manifold is abandoned so that now $D$ is a generalized Dirac operator acting on a Hilbert space $\mathcal{H}$, and $\mathcal{A}$ then is an algebra with an action on 
$\mathcal{H}$, which in general is noncommutative. 

The new fact in the Higgs case is that in addition to the standard spectral triple, denoted $(\mathcal{A}^{}_M,\mathcal{H}^{}_M,D^{}_M)$ now, one needs a further one to describe the Higgs sector, denoted $(\mathcal{A}^{}_F,\mathcal{H}^{}_F,D^{}_F)$, where the subscript 
$F$ signifies that the  algebra $\mathcal{A}^{}_F$ may be viewed as coming from a finite virtual set consisting of two points. Then one must define a product of two spectral triples, and applying the SAP to the product $M\times F$ one obtains  ($\left[11\right]$, see also 
$\left[22\right]$) the entire action with the Higgs field included, and the prefactor of its quadratic term shows the essential minus sign signalizing spontaneous symmetry breaking.  In addition, the derivation instructs that the Higgs field receives the status of a gauge field - originating from a discrete structure! 
Hence we see that the spectral action principle predicts essential physics; indeed, it is a principle of unprecedented predictive facilities because the gauge field action is solely determined by the Dirac operator by means of a definite calculational procedure. 
 
To mention a last point, another fact also to  be taken note of is that in the context of quintessence models the inclusion of the dilaton, which in the Jordan-Brans-Dicke theory is hampered by several stumbling blocks, e.g. that the kinetic term appears with the wrong sign, works perfectly in the noncommutative setting $\left[22,23\right]$.

Returning to the case of interest, since we know how to couple torsion to the Majorana fermions of dark matter  (see eq. \eqref{2.13}), we expect that the spectral action will tell us the relevant physics concerning the gauge field action for torsion. 

The necessary computations are rather intricate, but we can profit from the fact that one can rely on more recent literature 
$\left[24\right]$ since in the NCG context the question of the relevance of torsion in physics has found renewed interest 
$\left[22\right]$. In order to include the (scalar) Higgs field $\phi$ we make use of the results in $\left[25\right]$, from which  we extract the additional contribution of $\phi$, and so the final result for the total (minkowskian) action turns out to be
\begin{equation}\label{3.3}              
S^{}_{DM}=\int dx\,|e|\Big(\bar{\chi}(\overset{\circ}{D}-M)\chi\,+ \frac{1}{4}\,T^{\,5}{}_{\alpha}J_{\,5}{}^{\alpha}\Big)\,+
\end{equation}
$$+\,\int dx\,|e|\Big(\frac{1}{2\kappa^2}\,\big(2\Lambda_c-\overset{\circ}{R}\big)\,
-\frac{1}{4}\,T^{\,5}{}_{\alpha\beta}T_{\,5}{}^{\alpha\beta}+\frac{1}{2}\,m_5{}^2\,T^{\,5}{}_{\alpha}T_{\,5}{}^{\alpha}+
\big(\frac{1}{6}\overset{\circ}{R}-g\,T^{\,5}{}_{\alpha}T_{\,5}{}^{\alpha}\big)\left|\phi\right|^2\Big)$$
where $J^{\,5}{}_{\alpha}=i\bar{\chi}\gamma^{\,5}_{}\,\gamma_{\alpha}^{}\chi$ denotes the axial current of the DM Majorana fermions, 
$\Lambda_c$ the cosmological constant, $\kappa^{\,2}= 8\pi G$, and  
$T^{\,5}{}_{\mu\nu}=\partial^{}_{\mu}\,T^{\,5}{}_{\nu}-\partial^{}_{\nu}\,T^{\,5}{}_{\mu}$ the axial field strengths. In order to arrive at the final form \eqref{3.3}, the cutoff 
$\Lambda$ was hidden in the various coupling constants, and some field rescalings were performed. We have also included the gravitational contributions 
in the gauge field action, omitting the higher derivatives terms from the square of the Weyl tensor; a curl on top of a quantity signifies zero torsion. 

Altogether the final result for the action of the mediator field looks fairly simple; nevertheless, it dictates the relevant physics. On the one hand the SAP  of NCG  prognosticates that torsion is a dynamical degree of freedom. On the other hand torsion is also predicted to be a massive axial vector field since the mass term carries the correct sign in front. Beyond that, it is seen to be extremely massive since the torsion mass $m^{}_5$ is of similar order as the dark matter mass $M$; we even assume
\begin{equation}\label{3.4}              
m^{}_5 \gg M
\end{equation}
in analogy to the SM so that the axial vector field torsion can be viewed  as a superheavy companion of the $W$ and  $Z$ bosons. The condition \eqref{3.4} should also be needed to avoid naturalness resp. hierarchy problems.  Another remarkable prediction of the SAP is the very last term in the gauge field action, which couples torsion through the Higgs field to the SM; hence, the field mediating the novel gravitational forces between dark matter particles is coupled to the field which endows the constituents of the standard model with their corresponding masses. In what follows the construct introduced in the first two sections together with the action \eqref{3.3} is called the dark-matter-torsion (DMT) theory 

The result \eqref{3.3} is not entirely new; similar constructs, tentatively having been written down with rather different intentions, can be found in the literature $\left[26\right]$.

For a first orientation, our interest focuses on cold DM  (CDM) at cosmic time today; then the kinetic term of torsion is only slowly varying, it can thus be neglected against the dominant mass term, and so torsion can be integrated out to give
\begin{equation}\label{3.5}              
S^{}_{CDM}=\int dx\,|e|\Big(\frac{1}{2\kappa^2}\,\big(2\Lambda_c-\overset{\circ}{R}\big)\,+
\bar{\chi}(\overset{\circ}{D}-M)\chi\,- \frac{1}{32\,m_5{}^2}\,J^{\,5}{}_{\alpha}J_{\,5}{}^{\alpha}
-\frac{g}{16\,m_5{}^4}\,J^{\,5}{}_{\alpha}J_{\,5}{}^{\alpha}\,\left|\phi\right|^2\Big)
\end{equation}
$$J^{\,5}{}_{\alpha}=i\bar{\chi}\gamma^{\,5}_{}\,\gamma_{\alpha}^{}\chi.$$
Discarding the first and the last term, we thus see that this CDM-reduction of the DM-torsion system yields the Heisenberg-Weyl-Sciama-Kibble 
4-fermion contact interaction; due to the prefactor, is seen to be extremely weak. A similar result is known from the EC theory 
($\left[16\right]$, see also 
$\left[19\right]$). But the CDMT limit of slowly varying torsion fields is more general than ECT; only for $m_5{}^2=1/6\kappa_{}^2$ they are identical. In this way the ECT attains its status, as a special case of the DMT edifice. 

Of course, the instantaneous DM self-interaction is of nonrenormalizable type since \eqref{3.5} is an effective theory 
by construction, with \eqref{3.3} its UV completion. Comparing  \eqref{3.5} with the Bardeen-Cooper-Schrieffer theory of superconductivity one expects, disregarding the coupling term to the Higgs field, that the direct coupling between the DM particles is always attractive, independent of any spin orientation (cf. however $\left[27\right]$). 

As a last point, torsion as the mediator for DM interactions couples naturally also to the fermions of the SM, in particular to the quarks, so that the portal 
\begin{equation}\label{3.6}
\frac{(\bar{q}\,\gamma^{\,5}_{}\gamma^{}_{\mu}\,q)(\bar{\chi}\,\gamma_{\,5}^{}\gamma_{}^{\mu}\,\chi)}{4m_5{}^2}
\end{equation} 
is opened, being suppressed by two powers of a 
superheavy mass; it is intensively studied in the literature $\left[1\right]$. Another channel is available through the Higgs portal 
$\left[1,12\right]$) that couples the Higgs  to torsion. 

\section{Dark-matter-torsion theory and cosmology}

Returning to the introduction, there we have reviewed work done on Wimpzilla DM, which shows that such DM particles can sufficiently be produced at the end of inflation through freeze-in to generate (at least part of) the measured DM abundance. It remains to investigate the impact of the  additional ingredients from the dark-matter-torsion theory of the preceding section, in particular the impact of the torsion field. 
So we must confirm that nothing serious can happen during the cosmological evolution since the end of inflation till today. This amounts to ascertaining whether the Friedmann equations of cosmology remain valid or are somehow altered if torsion is included as the missing gravitational mediator.  

What comes to mind is to resort to the Weyssenhoff concept of a perfect spin fluid $\left[28\right]$ in the framework of EC theory. This approach was $\left[29\right]$ and still is $\left[30\right]$ intensively investigated in the literature, It was brought to considerable perfection in the work of Obukhov and Karotky (see $\left[30\right]$)\footnote{Nevertheless, there is a point which we dispute. In their work they claim that there is a theorem, guaranteeing that the non-symmetric energy momentum tensor of the spin fluid can be written in the form $P^{\mu}{}{}_{\nu}=u^{\mu}P_{\nu}$ where $u$ is the velocity field of the fluid; it is a crucial point to be reconsidered since in the expansion in terms of moments of Mathisson, Papapetrou, Dixon and others this property only holds in zeroth order of the expansion in terms of moments; it is no longer valid in the first order of the expansion in terms of moments, being needed for the treatment of the spin.}, 
with the outcome that the only modification in the Friedmann equations consists in the replacement of the energy density and the pressure by
$\rho\to\rho^{\text{eff}}=\rho -\aleph s^2$ and $p\to p^{\text{eff}}=p -\aleph s^2$ where $\aleph$ is a positive constant and $s^2$ the averaged value of the spin tensor squared. Apart from the fact that the numerical value of $\aleph$ is not known, the result does not look plausible since one expects that the torsion field should increase the energy density instead of reducing it. To say the least, the Weyssenhoff spin fluid does not seem to describe an adequate phenomenological model. Perhaps one weak point is that the Weyssenhoff form $S^{\mu}{}_{\nu\rho}=u^{\mu}_{}S^{}_{\nu\rho}$ of the spin tensor is not valid for a Dirac fermion, for which 
$S_{}^{\mu\nu\rho}$ is totally skew, and another one that no restrictions on torsion are imposed.  

As the Weyssenhoff spinning fluid should probably not be a viable model, for whatever reason, we follow another route, having been founded by Tsamparlis ($\left[31\right]$, see also $\left[32\right]$), and put further in $\left[33\right]$ under various different aspects. Since all these works refer to the EC theory, we must adapt the setting to the present case. 

In order to generalize cosmology to the realm of Riemann-Cartan spaces. the basic step consists in determining those torsion configurations, which are compatible with the cosmological principle. Thus, if $L^{}_{\xi}$ denotes the Lie derivative along a vector field $\xi$, in a Riemann-Cartan space the two conditions
\begin{equation}\label{4.1}
L^{}_{\xi}\,g^{}_{\mu\nu}=0\qquad\qquad\qquad L^{}_{\xi}\,T^{}_{\mu\nu\rho}=0
\end{equation}
must then be obeyed, where $\xi$ runs through a basis of vector fields of the group $T(3)\rtimes O(3)$; for the present, torsion is not restricted somehow. The first of these two invariance conditions concerning the metric tensor is the standard one, leading to the Friedmann-Lema\^{i}tre-Robertson-Walker metric. As to the second invariance condition imposed on the torsion tensor, applying tools elaborated in $\left[34\right]$ it is straightforward to show that there are two solutions
\begin{equation}\label{4.2}
{_{}^{\text{V}}T_{}^{}}_{0j}{}^k=\,F\,\delta_j{}^k\qquad\qquad\qquad {_{}^{\text{A}}T_{}^{}}_{ij}{}^k=
\,f\,\varepsilon_{\,ij}{}^k
\end{equation}
with the prefactors $F$ and $f$ depending on cosmic time; all torsion components not shown vanish, the superscripts V and A stand for vector and axial vector. Our case of interest is CDMT theory so that only the axial solution is allowed; in the comoving frame, with $u$ the velocity field of DM, it reads
\begin{equation}\label{4.3}
T^{}_{\,\alpha\beta\gamma}=
(f/3m^{}_5\kappa)\,\varepsilon_{\,\alpha\beta\gamma\delta}\,u_{}^{\delta}
\end{equation}
with the prefactor chosen for convenience. To proceed, we return to the CDM reduction of the action \eqref{3.3}, but only keep the essential terms:
\begin{equation}\label{4.4}              
S^{}_{CDM}=\int dx\,\sqrt{-|g|}\Big(\bar{\chi}(\overset{\circ}{D}-M)\chi\,+ \frac{1}{4}\,T^{\,5}{}_{\alpha}J_{\,5}{}^{\alpha}
-\frac{1}{2\kappa^2}\,\overset{\circ}{R}
+\frac{1}{2}\,m_5{}^2\,T^{\,5}{}_{\alpha}T_{\,5}{}^{\alpha}\Big).
\end{equation}
Using $J_{\,5}{}^{\alpha}=-4m_5{}^2\,T_{\,5}{}^{\alpha}$ we have
\begin{align}\label{4.5}
\begin{split}
S^{}_{CDM}&=\int dx\,\sqrt{-|g|}\Big(-\frac{1}{2\kappa^2}\overset{\,\circ}{R}+\bar{\chi}(\overset{\circ}{D}-M)\chi\,
-\frac{1}{2}\,m_5{}^2\,T^{\,5}{}_{\alpha}T_{\,5}{}^{\alpha}\Big)\\
&=-\frac{1}{2\kappa^2}\,\int dx\,\sqrt{-|g|}\Big(\overset{\circ}{R}+f_{}^2g^{}_{\mu\nu}\,u_{}^{\mu}\,u_{}^{\nu}\Big)+
\,\int dx\,\sqrt{-|g|}\bar{\chi}(\overset{\circ}{D}-M)\chi
\end{split}
\end{align}
and varying the metric gives
\begin{equation}\label{4.6}
\overset{\scriptscriptstyle\circ}{R}{}^{}_{\mu\nu}-
\frac{1}{2}\,g^{}_{\mu\nu}\,\overset{\scriptscriptstyle\circ}{R}=\kappa_{}^2\big(T{}^{}_{\mu\nu}+
f_{}^2(u^{}_{\mu}\,u^{}_{\nu}+\frac{1}{2}\,\,g^{}_{\mu\nu})\big)
\end{equation}
with $T^{}_{\mu\nu}$ the symmetric energy-momentum tensor of matter. In the context of cosmology it takes the canonical form of the 
energy-momentum tensor of an ideal fluid
\begin{equation}\label{4.7}
T^{}_{\mu\nu}=(\rho+p)u^{}_{\mu}u^{}_{\nu}-p\,g^{}_{\mu\nu}
\end{equation}
with $\rho$ the mass density and $p$ the pressure. As \eqref{4.6} instructs, in the presence of torsion it gets modified; introducing
\begin{equation}\label{4.8}
\tilde{T}{}^{}_{\mu\nu}=T{}^{}_{\mu\nu}+
f_{}^2(u^{}_{\mu}\,u^{}_{\nu}+\frac{1}{2}\,\,g^{}_{\mu\nu})
\end{equation}
we see that for $\tilde{T}{}^{}_{\mu\nu}$ the form is preserved since
\begin{equation}\label{4.9}
\tilde{\rho}=\rho\,+\,\frac{3}{2}\,f^2\qquad\qquad\qquad \tilde{p}=p\,-\,\frac{1}{2}\,f^2.
\end{equation}
Accordingly, the presence of torsion increases the mass density and diminishes the pressure. It is indeed the result we want; beyond that, torsion leaves the Friedmann-Lema\^{i}tre equations of cosmology intact. For the special case of the EC theory a 
similar result is known to hold.\footnote{For a comment on troubles  in the literature, concerning signs and prefactors of the result \eqref{4.9}, see the note in Sch\"{u}cker \& Tilquin $\left[33\right]$.} The different signs in \eqref{4.9} are essential; because torsion as a bosonic field enhances the energy density, and because cold dark matter is almost pressureless. 

For the present 
dark-matter-torsion theory the result says, if taken literally, that the dark-matter abundance at  the end of inflation is raised through the subsequent cosmic evolution to its value today, and the pressure of dark matter is lowered in this process to a value near zero 

\section{Summary and prospect}

To give a resume, we have collected diverse facts to provide some evidence that it should make sense to consider the dark-matter-torsion theory, as embodied in eq. \eqref{3.3}, with superheavy right-handed Majorana fermions constituting dark matter and supermassive axial torsion the mediating field, as a promising candidate to resolve the dark matter problem. The proposed solution requires a natural minimal extension of both the standard model and general relativity, which in the limit of a slowly varying mediator field, i.e. in the CDM limit of non-dynamical torsion, becomes a modification of the Einstein-Cartan theory. 

Of course, the proposal requires experimental certification. As the parameter space is rather large, means of a reduction are needed. Since the SAP acts at a scale of the order of the unification scale, one way could be to process the parameters of the DMT theory through the renormalization group equations to obtain an estimate of the present values. Another task that remains is to extend the above analyses to the entire SM, in particular to take the triple of superheavy Majorana fermions into account, one for each family, requiring a natural extension of the dark matter sector.

\end{document}